\documentclass[notitlepage,superscriptaddress,showpacs,nobalancelastpage,twocolumn,aps,longbibliography,prl]{revtex4-2}

\usepackage[english]{babel}
\RequirePackage[T1]{fontenc}
\RequirePackage{times} 

\usepackage{siunitx} 
\usepackage[italicdiff]{physics}
\usepackage{amsfonts}
\usepackage{lipsum}
\usepackage{multirow}
\usepackage{subfigure}
\usepackage{amsmath}
\usepackage{amssymb}
\usepackage{graphicx,epstopdf}
\usepackage{xspace}
\usepackage{array}
\usepackage[hidelinks]{hyperref}
\usepackage[dvipsnames]{xcolor}
\usepackage{my_symbols}
\usepackage{CJK}
\usepackage{bm}
\usepackage{verbatim}
\usepackage{tikz}
\usepackage{comment}
\usepackage{marginnote}
\usepackage[textwidth=1.5cm]{todonotes}
\usetikzlibrary{calc,3d}
\usetikzlibrary{arrows}
\usepackage[normalem]{ulem}
\usepackage{soul}
\usepackage{dsfont}

\sethlcolor{green}

{\par}

\newcounter{mycomment}

\includecomment{comment} 

\begin{document}

\begin{CJK*}{UTF8}{gbsn} 
\title{Mechanism for the Broadened Linewidth in Antiferromagnetic Resonance}
\author{Yutian Wang}
\affiliation{Department of Physics and State Key Laboratory of Surface Physics, Fudan University, Shanghai 200433, China}
\author{Jiang Xiao (萧江)}
\email[Corresponding author:~]{xiaojiang@fudan.edu.cn}
\affiliation{Department of Physics and State Key Laboratory of Surface Physics, Fudan University, Shanghai 200433, China}
\affiliation{Institute for Nanoelectronics Devices and Quantum Computing, Fudan University, Shanghai 200433, China}
\affiliation{Shanghai Research Center for Quantum Sciences, Shanghai 201315, China}
\affiliation{Shanghai Branch, Hefei National Laboratory, Shanghai 201315, China}

\begin{abstract}
The linewidth of antiferromagnetic resonance (AFMR) is found to be significantly broader than that of ferromagnetic resonance (FMR), even when the intrinsic Gilbert damping parameter is the same for both systems. We investigate the origin of this enhanced damping rate in AFMR by studying a bipartite magnet model. Through analytical calculations and numerical simulations, we present three perspectives on understanding this linewidth broadening in AFMR: i) The non-dissipative Heisenberg exchange interaction develops a damping-like component in the presence of Gilbert damping, ii) The transverse component of the exchange coupling reduces the AFMR frequency, thereby increasing the damping rate, and iii) The antiferromagnetic eigenmode exhibits characteristics of a two-mode squeezed state, which is inherently linked to an enhanced damping rate. Our findings provide a comprehensive understanding of the complex dynamics governing magnetic dissipation in antiferromagnet and offer insights into the experimentally observed broadened linewidths in AFMR spectra.
\end{abstract}
\maketitle
\end{CJK*}

\emph{Introduction -} 
Ferromagnetic resonance (FMR) refers to the precession of magnetic moments in a ferromagnetic material around an external magnetic field at a specific resonance frequency \cite{kittel_theory_1948,griffiths_anomalous_1946}. This phenomenon is widely used to study magnetic properties and has various applications, including magnetic storage, spintronics, and magnetic resonance imaging \cite{Michael_Farle_1998, hillebrands_spin_2002, hillebrands_spin_2003, hillebrands_spin_2006}. The linewidth of FMR, which characterizes the rate at which the magnetization returns to equilibrium after being perturbed, is affected by several factors, with magnetic damping being a significant contributor. Therefore, the linewidth of FMR provides valuable information about the magnetic properties of materials, particularly regarding the investigation and characterization of magnetic damping mechanisms.

Just like ferromagnetic materials exhibit ferromagnetic resonance (FMR), antiferromagnetic materials demonstrate antiferromagnetic resonance (AFMR) \cite{kittel_theory_1951, keffer_theory_1952}. The study of AFMR also offers insights into the dynamics and properties of antiferromagnetic materials \cite{johnson_antiferromagnetic_1959, ohlmann_antiferromagnetic_1961, mori_antiferromagnetic_1962, foner_high-field_1963, gomonay_spintronics_2014,jungwirth_antiferromagnetic_2016,gomonay_concepts_2017, baltz_antiferromagnetic_2018, nemec_antiferromagnetic_2018,gomonay_antiferromagnetic_2018}. 
The resonance frequency of AFMR is influenced by factors such as the strength of the exchange interaction, magnetic anisotropy, and the applied field. Both the AFMR resonance frequency and its linewidth provide valuable information about antiferromagnetic order, spin wave excitations, magnetic anisotropies, and especially the spin-related interactions in antiferromagnetic materials.

The linewidth of the resonance is typically proportional to the resonance frequency. The damping rate, the ratio between the linewidth and the resonance frequency, roughly characterizes how many cycles the oscillation can perform before damping out.
This damping rate is usually determined by a phenomenological damping coefficient, such as the viscous coefficient in a driven mechanical oscillator or the Gilbert damping parameter in FMR. Mechanisms like spin pumping give a correction to the damping coefficient, enhancing the Gilbert damping parameter \cite{tserkovnyak_enhanced_2002, tserkovnyak_spin_2002, tserkovnyak_nonlocal_2005, cheng_spin_2014, moriyama_enhanced_2020}. There are some other mechanisms, such as two-magnon scattering \cite{zakeri_spin_2007}, that will introduce an extra damping effect aside from Gilbert damping. In coupled systems, the damping rate of the normal modes does not typically exceed the damping rate of the individual subsystems, provided that the coupling is coherent and does not introduce additional dissipation \cite{azzawi_magnetic_2017}.

The antiferromagnet can be conceptualized as two magnetic sublattices interconnected through Heisenberg exchange coupling. In this context, it is not surprising that the antiferromagnet is likened to a pair of coupled oscillators, with each magnetic sublattice represented as an oscillator. From this viewpoint, the damping rate for the antiferromagnet is expected to be similar to that of coupled oscillators. However, this paper highlights that this seemingly intuitive perspective is actually incorrect. The distinction between coupled oscillators and the antiferromagnet lies in the fact that coupled oscillators couple two normal particles, while the antiferromagnet, in the mean-field approximation, couples a particle with its anti-particle \cite{wu_landau_2001, wu_superfluidity_2003}. 
This dissimilarity also results in distinct characteristics in the linewidth of the spectrum in the antiferromagnet: the damping rate in the antiferromagnet can exceed the damping rate of individual sublattices. 

\begin{figure*}[ht]
    \centering
    \includegraphics[width=\textwidth]{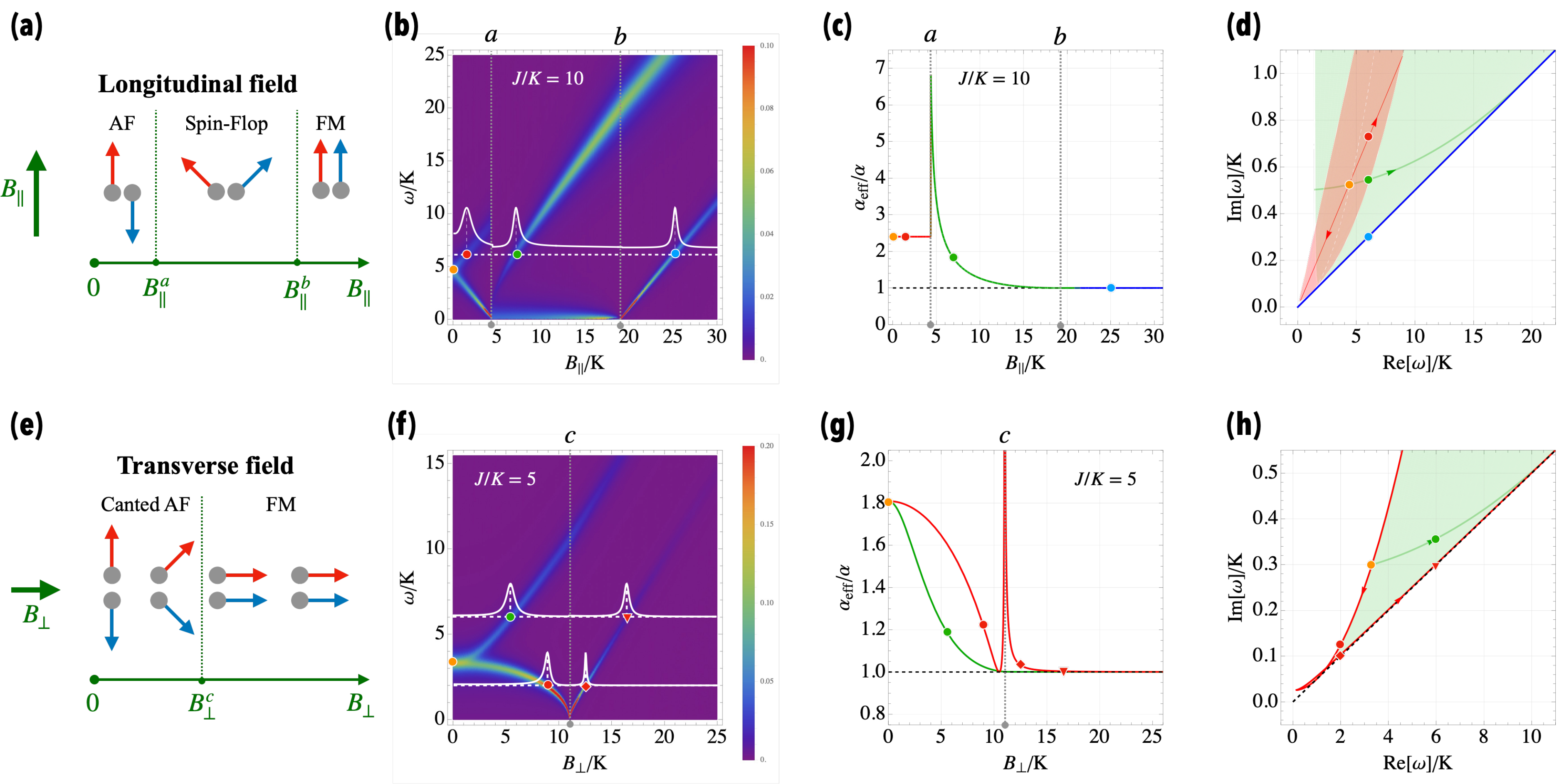}
    \caption{(a, e) The longitudinal and transverse magnetic field modify the ground state of a bipartite magnet. 
    (b, d) The simulated spin wave excitation in a bipartite magnet with increasing longitudinal and transverse field.
    (c, g) The damping enhancement ratio $\alpha_\ssf{eff}/\alpha$ as function of field for the resonances in (b, d).
    (d, h) The calculated spin wave eigenfrequencies upon various ground states in a complex-$\omega$ plane with varying $J$ and $B$. The damping rate for the AF, spin-flip, and canted AF state is lower bounded by the FM damping rate $\alpha$. }
   \label{fig:long_trans}
\end{figure*}

The dissipation is built-in in terms of the Gilbert damping parameter $\alpha$, which has been shown to be a more realizable parametrization of the magnteic dissipation\cite{gilbert_classics_2004,kambersky_spin-wave_1975}. An experimentally more relevant parameter for characterizing magnetic dissipation is the linewidth, which is directly related to the imaginary part of the eigenfreuqency $\Im{\omega}$. Apart from a field/frequency-independent contribution, the linewidth is typically found to be proportional to the resonance frequency, \ie $\Im{\omega} \propto \Re{\omega}$. We define a damping rate as the ratio between the imaginary and the real part of the eigenfrequency:
\begin{equation}
    \alpha_\ssf{eff} \equiv \frac{\Im{\omega}}{\Re{\omega}},
\end{equation} 
whose inverse characterizes the number of cycles before an excitation damping out. In FMR experiments, this damping rate is usually associated with the Gilbert damping constant used in the LLG equation above: $\alpha_\ssf{eff} = \alpha$. Naively, this identification makes sense because the Gilbert damping is the only dissipative mechanism in the exchange coupled LLG equations in \Eq{eqn:LLG}, and the Heisenberg coupling does not introduce any extra dissipation. In this Letter, we emphasize that the damping rate for AF actually is larger than the sublattice Gilbert damping $\alpha$, and more importantly, we provide an explaination for the linewidth broadening in AFM in three different perspectives, from the microscopic torque analysis to a quasiparticle point view, then connecting to the generalized concept of spin wave polarization in antiferromagnet.


\emph{Bipartite Model - }
We consider a minimal model of a bipartite magnets described by the following Hamiltonian 
\begin{equation}
    H = 
    - \sum_{j=1}^2 \qty[\frac{K}{2} \qty(\mb_j\cdot\hbz)^2 + \bB \cdot \mb_j ]
    + J \mb_1\cdot\mb_2,
\end{equation}
where $K, J$, and $\bB$ represent the uniaxial anisotropy along $\hbz$, the Heisenberg exchange, and the external magnetic field, respectively. 
The dynamics of the $\mb_{1,2}$ are governed by the phenomenological Landau-Lifshitz-Gilbert equation \cite{landau_theory_1992,gilbert_classics_2004,keffer_theory_1952} 
\begin{subequations}
    \label{eqn:LLG}
    \begin{alignat}{3}
    &\dot{\mb}_1 &=-\gamma \mb_1\times(\bB + K m_1^z\hbz-J\mb_2)
    &+\alpha \mb_1\times \dot{\mb}_1 \\
    &\dot{\mb}_2 &=-\gamma \mb_1\times(\bB + K m_2^z\hbz-J\mb_1)
    &+\alpha \mb_1\times \dot{\mb}_2.
    \end{alignat}
\end{subequations}
The Heisenberg exchange term favors the (anti)ferromagnetic configuration for negative (positive) $J$.
In the absence of external magnetic field, the resonance frequencies for the bipartite magnet are given by 
\begin{subequations}
    \label{eqn:frequency_nosf}
    \begin{align}
    \omega_\pm^{\rm FM} &= \qty(K+\abs{J}\pm\abs{J})(1+i\alpha), \\
    \omega_\pm^{\rm AF} &= 
    \pm\sqrt{(\abs{J}+K)^2-\abs{J}^2} 
    + i\alpha(\abs{J}+K). 
    \end{align}
\end{subequations}
The opposite signs of eigenfrequencies $\Re{\omega_\pm^{\rm AF}}$ for the AF phase indicate their opposite polarizations, being right-handed and left-handed, respectively \cite{keffer_theory_1952}. 
From these expressions, one immediately sees that the damping rate for FM is equal to the sublattice Gilbert damping constant $\alpha$, while the damping rate for AF is larger than $\alpha$:
\begin{equation}
\label{eqn:alphaeffAF}
  \frac{\alpha_\ssf{eff}^{\rm AF}}{\alpha} = \frac{1}{\sqrt{1-\qty[\abs{J}/(\abs{J}+K)]^2}} = \cosh(2r) \ge 1
\end{equation}
with $\tanh(2r) = \abs{J}/(\abs{J}+K)$.
We should note that this enhanced damping for the eigenmodes does not affect the fluctuation-dissipation theorem, which is governed by the imaginary part of the linear response susceptibility function.
Several literatures \cite{gurevich_magnetization_1996, xu_ferromagnetic_2018, kamra_gilbert_2018, suto_microwave-magnetic-field-induced_2019, miwa_giant_2021} have touched on this damping rate enhancement in AF, however the physical explanation and intuitive understanding on the enhancement remain elusive. Therefore, in this paper, we try to demonstrate the phenomenon based on a theoretically minimum bipartite model and provide the phyiscal understanding for the damping rate enhancement in AF. 

\emph{Simulation -}
To illustrate the enhanced damping in AF, considering a bipartite magnet initially in an antiferromagnetic state with a positive exchange constant ($J > 0$), the application of an external field $\bB$ can induce a ground state alteration. This process is scrutinized under two distinct conditions: when the field is aligned parallel ($\bB \| \hz$) and perpendicular ($\bB \perp \hz$) to the easy axis $\hz$ as shown in \Figure{fig:long_trans}(a,e).

\Figure{fig:long_trans}(b) shows the simulated spin wave excitation spectrum with increasing longitudinal field $B_\|$ for $J/K = 10$. The system experiences a sequential transformation from antiferromagnetic ground state through a spin-flop state into a ferromagnetic state at critical fields denoted as $B_\|^a = \sqrt{K(2J-K)}$ and $B_\|^b = 2J-K$ \cite{yung-li_spin_1964, hu_energetics_2023}. Here we focus on the linewidth of the resonances for these three phases. A line cut at frequency $\omega = 6K$ has three reasonance peaks, in the AF, spin-flop, and ferromagnetic phases, respectively. Each peak has a  different linewidth, even though it is for the same system with fixed Gilbert damping of $\alpha = 0.05$ at the same resonance frequency. 
\Figure{fig:long_trans}(c) shows the damping rate enhancement for the modes in \Figure{fig:long_trans}(b).
\Figure{fig:long_trans}(d) shows the calculated eigenfrequencies (see Appendix A) for these three phases on a complex-$\omega$ plane, \cite{yung-li_spin_1964, gurevich_magnetization_1996}. The FMR is right on the straight line of slope $\alpha$, while the excitation upon the AF and spin-flop state have a higher damping rate than $\alpha$. 

\Figure{fig:long_trans}(f) shows a similar simulated spin wave excitation spectrum at $J/K = 5$ with the magnetic field transverse to the anisotropy axis. 
In this case, the ground state evolves from antiferromagnetic to a canted configuration, eventually achieving a ferromagnetic state at a critical field $B_\perp^c = 2J+K$. At a line cut at fixed frequency ($\omega = 2K, 6K$), there are two resonance peaks for canted antiferromagnetic and ferromagnetic ground states, respectively. It can be seen that the linewidth in the canted AF phase is larger than that in the FM phase. 
\Figure{fig:long_trans}(g) shows the damping rate enhancement for the modes in \Figure{fig:long_trans}(f).
\Figure{fig:long_trans}(h) shows the calculated complex resonance frequency on a complex-$\omega$ plane (see Appendix B), which is above the slope $\alpha$ as well. Therefore, in the above examples for the longitudinal and transverse field cases in \Figure{fig:long_trans}, we see that the damping rate is indeed enhanced for the antiferromagnet, but also for the non-collinear spin-flop or canted antiferromagnet.  

To understand the mechanisms behind enhancing the damping rate, especially in the antiferromagnet, we employ three distinct analytical approaches. Firstly, a microscopic examination of how the non-dissipative exchange torque contributes to damping. Secondly, an analysis that distinguishes the effects of longitudinal and transverse components of the Heisenberg exchange interactions. Lastly, attention is focused on the influence of magnon squeezing and polarization in modulating the damping rate. Each perspective offers unique insights into the complex dynamics governing magnetic damping in antiferromagnetic materials.

\begin{figure}[t]
    \centering
    \includegraphics[width=\columnwidth]{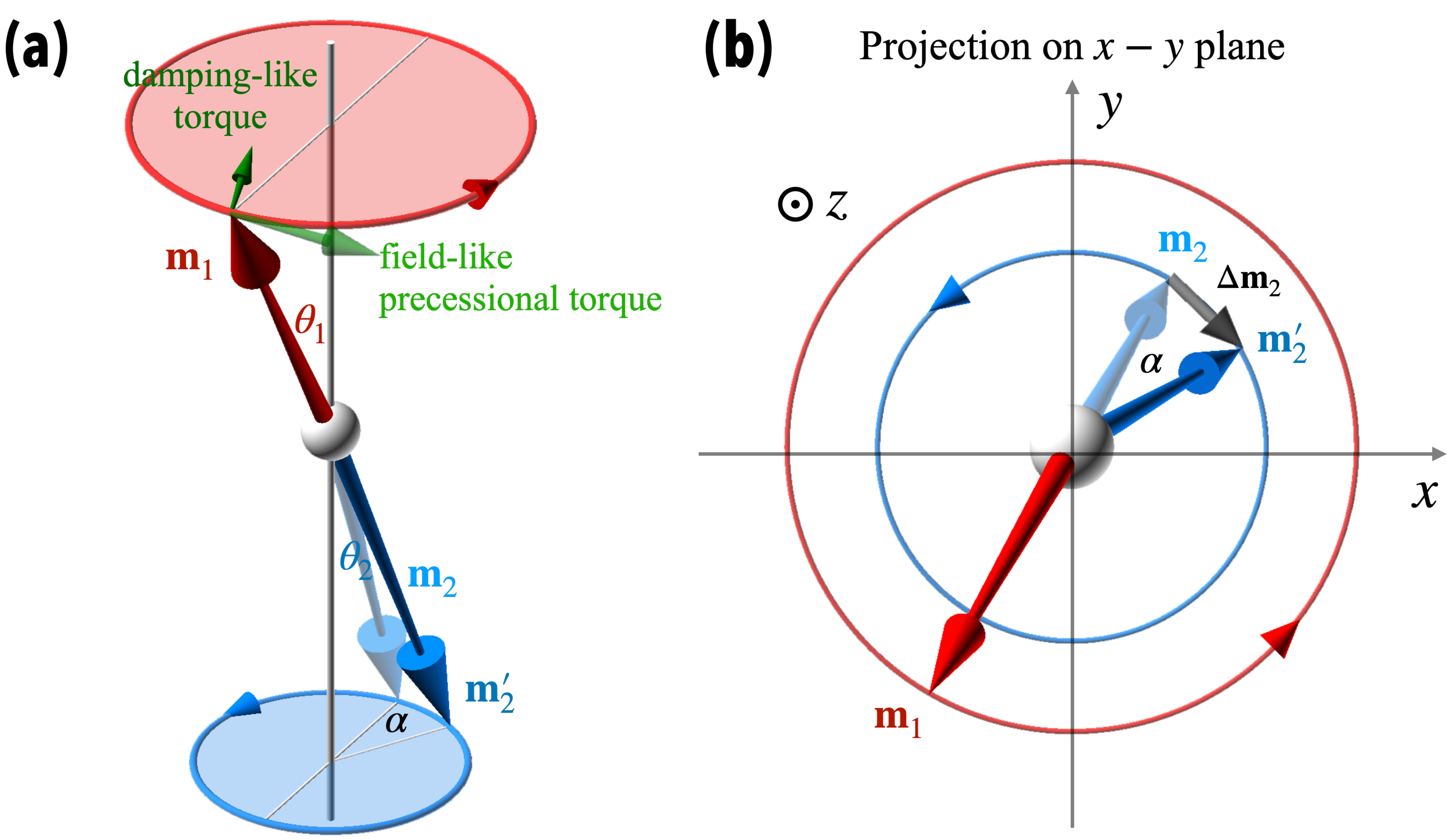}
    \caption{Comparison of the antiferromagnetic resonance state without and with Gilbert damping in a 3D view (a) and projection on the $x-$y plane (b). When $\alpha = 0$, $\mb_1$ and $\mb_2$ point in opposite direction in the projection view. When $\alpha \neq 0$, $\mb'_2$ has an extra phase delay of $\alpha$. For this damping modulated state, the exchange torque $J\mb_1\times\mb'_2$ has a damping-like component of magnitude $\abs{J\mb_1\times\Delta\mb_2} = J\alpha\theta_2$. }
    \label{fig:torque}
\end{figure} 

\emph{Torque analysis -}
The Heisenberg exchange coupling is inherently even in time reversal and thus non-dissipative by itself. However, the non-dissipative exchange torque can influence the damping behavior with a time-reversal broken antiferromagnetic state. For an antiferromagnetic eigen mode of circular polarization, the magnetic moments $\mathbf{m}_{1,2}$ undergo circular trajectories at frequency $\Omega = \sqrt{K(2\abs{J}+K)}$ around the axis $\hat{\mathbf{z}}$ with respective cone angles $\theta_{1,2}$ (see \Figure{fig:torque}). For $\mb_1$, the field-like precessional torque is $\abs{\dot{\mb}_1} = \Omega\theta_1$, and the damping-like Gilbert torque $\abs{\alpha \mathbf{m}_1 \times \dot{\mathbf{m}_1}} \simeq \alpha \Omega \theta_1$. Therefore, if only Gilbert torque contributes to the damping, the damping rate would erroneously appear as constant $\alpha$, not the enhanced value of $\alpha_\ssf{eff}^\ssf{AF}$ in \Eq{eqn:alphaeffAF}. 

The mistake in the above analysis lies in the assumption that the exchange torque $J \mb_1\times\mb_2$ points perfectly in the $x$-$y$ plane, thus is a purely field-like precessional torque. This is indeed the case if $\mb_1, \mb_2$, and $\hbz$ are all in the same plane, or equivalently $\mb_1$ and $\mb_2$ point in opposite directions in the $x$-$y$ projection plane (see \Figure{fig:torque}). However, when taking the magnetic damping into account, an important observation for an antiferomangetic dynamical state is that the three vectors $\mb_{1,2}(t)$ and $\hbz$ do not lie in the same plane. Instead, $\mb_2$ becomes $\mb'_2$ has an extra (average) phase delay of $\alpha$ relative to $\mb_1$ in the projection plane in addition to the original $\pi$ phase, as shown in \Figure{fig:torque}(b) (see Appendix C). 
This misalignment between $\mb_1$ and $\mb_2$ leads to a tilting of the exchange torque out of the precessional plane, thus gives rise to a damping-like component of magnitude $J(\alpha\theta_2)$ on $\mb_1$. 
Consequently, the total damping-like torque on $\mb_1$ now reads
\begin{equation}
    \alpha \Omega \theta_1 + \alpha J\theta_2
    = \alpha (K+J) \theta_1
    = \frac{K+J}{\sqrt{K(2J+K)}} \alpha\Omega\theta_1,
\end{equation}
where $\theta_2/\theta_1 = J/(K+J+\Omega)$ is used \cite{keffer_theory_1952}.
In comparison with the precessional torque $\Omega\theta_1$ on $\mb_1$, we recovered the enhanced damping rate as in \Eq{eqn:alphaeffAF}. A similar analysis applies on $\mb_2$ as well.

\emph{Longitudinal \& transverse coupling - }
An alternative understanding of the enhanced damping rate in AF is to rewrite the exchange coupling as:
\begin{align}
    \label{eqn:JJ}
    J\mb_1 \cdot \mb_2 
    = Jm_1^zm_2^z + J' \qty(m_1^xm_2^x + m_1^ym_2^y),
\end{align}
separating the longitudinal and transverse exchange coupling. For the typical isotropic Heisenberg interaction, $J' = J$. Consequently, the eigenfrequencies in \Eq{eqn:frequency_nosf} are rewritten as
\begin{subequations}
\begin{align}
    \label{eqn:wAFJJ}
    \omega_\pm^{\rm FM} &= (1+i\alpha)\qty(K+\abs{J}\pm \abs{J'}), \\
    \omega_\pm^{\rm AF} &= \pm \sqrt{(K+\abs{J})^2- \abs{J'}^2} +i\alpha(K+\abs{J}),
\end{align}
\end{subequations}
which indicate that the longitudinal and transverse coupling are qualitatively different. More interestingly, for AF case, the transverse coupling $J'$ does not affect the dissipative imaginary part, but reduces the real part of the eigenfrequencies. 
Consequently, the damping rate for AF spin wave becomes larger than $\alpha$. The reduction of the eigenfrequency due to $J'$ also implies the gap difference between the left- and right-circular AF modes $\omega_\pm^{\rm AF}$ reduces because of the transverse coupling, manifesting a level attraction behavior \cite{wang_antiferromagnetic_2021}. 
For comparison, in the FM case, the transverse coupling $J'$ affects both real and imaginary parts in the same fashion, thus leaving the damping rate unchanged. 

\emph{Two-mode Squeezing -} In terms of the magnon creation and annihilation operators for $\mb_{1,2}$, the Hamiltonian for the bipartite magnet can be written as
\begin{equation*}
    \label{eqn:H}
    \hH = (K+|J|) (\ha_1^\dagger\ha_1 + \ha_2^\dagger\ha_2) + |J'| \begin{cases}
        -(\ha_1^\dagger \ha_2 + \ha_1\ha_2^\dagger) \qfor \mbox{FM}\\
        \ha_1^\dagger \ha_2^\dagger + \ha_1\ha_2 \qfor \mbox{AF}
    \end{cases},
\end{equation*}
which shows that the transverse $J'$ coupling is a particle-number-conserved coupling for FM but particle-number-non-conserved coupling for AF.
Hamiltonian with the particle-number-non-conserved form is known to give rise to mode squeezing. In the present case, the transverse coupling in AF causes the two-mode squeezing between the excitations of $\mb_1$ and $\mb_2$ with squeezing parameter $r$: $\tanh(2r) = |J'|/(K+|J|)$ (see Appendix D). Surprisingly, this squeezing parameter $r$ is the same $r$ in the damping rate enhancement in \Eq{eqn:alphaeffAF}. This means that the damping rate enhancement in AF is related to the squeezing caused by the particle-non-conserved coupling. In contrast, the FM Hamiltonian has no squeezing, so there is no damping rate enhancement in FM. 

Similar to two-mode squeezing in the antiferromagnetic magnon discussed here, there can be single-mode squeezing for ferromagnetic magnon. The squeezing Hamiltonian can be found in ferromagnetic systems by anisotropy, inhomogeneous magnetic texture, or dipolar interactions. Such ferromagnetic squeezing also leads to the damping rate enhancement in ferromagnetic spin waves. Because the squeezing of ferromagnetic magnons also implies the elliptical polarization of spin wave, it is no surprise that damping rate enhancement is also found for non-circular ferromagnetic spin waves \cite{patton_linewidth_1968, kambersky_spin-wave_1975} and the soft modes in magnetic Skymions with inhomogeneous magnetic texture \cite{rozsa_effective_2018}.

\emph{Conclusion -}
Our results reveal that the dissipationless exchange interaction can significantly influence the dissipative properties of antiferromagnetic resonance, and more broadly, the spin excitations in systems with inhomogeneous magnetic ground states. This effect can be elucidated from both microscopic and macroscopic viewpoints. At the microscopic level, the intrinsic Gilbert damping slightly alters the antiferromagnetic mode, allowing the originally non-dissipative exchange torque to develop a damping-like component. At the macroscopic level, the dynamical (transverse) exchange interaction between the two magnetic sublattices serves to decrease the antiferromagnetic resonance frequency, thereby enhancing the damping rate relative to both the linewidth and the resonance frequency. Additionally, the increased damping in antiferromagnetic systems is linked to the fact that the AF eigenmode exhibits characteristics of a squeezed mode.

\emph{Acknowledgements -} 
J. X. acknowledges fruitful discussion with Gerrit Bauer. This work was supported by the National Key Research and Development Program of China (Grant no. 2022YFA1403300) and Shanghai Municipal Science and Technology Major Project (Grant No.2019SHZDZX01).

\bibliographystyle{apsrev4-2}
\bibliography{ref}

\appendix 

\section{Appendix A: Complex frequencies on longitudinal field scan}
For an antiferromagnetic bipartite magnet system, when the external field is applied parallel to the easy axis ($\bB \| \hz$), there are two phase transitions: from antiferromagnet (AF) to spin-flop (SF) at $B_\|^a = \sqrt{K(2J-K)}$, and from SF to ferromagnet at $B_\|^b = 2J-K$.
From the linearized coupled LLG equations, we can get the complex frequencies for the spin wave excitation in these three phases
\begin{subequations}
    \label{eqn:frequency}
    \begin{align}
    \omega_\pm^\ssf{AF} &= 
    \qty[\pm \Omega +i\alpha(J+K)]\qty(1\pm\frac{B}{\Omega}) 
    &,~& B \le B_\|^a, \\
    \omega_+^\ssf{SF}
    &= 2\sqrt{JJ_B} 
    +i\alpha (J+J_B)
    &,~& B_\|^a < B < B_\|^b, \\
    \omega_\pm^\ssf{FM} &= \qty(B+K-J\pm J)(1+i\alpha) 
    &,~& B \ge B_\|^b, 
    \end{align}
\end{subequations}
where $\Omega=\sqrt{K(2J+K)}$ is the antiferromagnet eigenfrequency and $2J_B = (2J+K)(B/B_\|^b)^2-K$. The opposite signs of eigenfrequencies $\Re{\omega_\pm^\ssf{AF}}$ for the AF phase indicate their opposite polarizations, being right-handed and left-handed, respectively. In SF phase, there is also a zero frequency Goldstone mode with $\omega_-^\ssf{SF} = 0$ precessing about $\hbz$.

\section{Appendix B: Complex frequencies on transverse field scan}
When the external field is applied perpendicular to the easy axis ($\bB \perp \hz$), there is only one phase transition at $B_\perp^c=2J+K$, separating canted antiferromagnet (CAF) phase from the and saturated ferromagnetic (FM) phase. The complex frequencies for CAF and FM phases are calculated as
\begin{subequations}
    \begin{align}
        \omega_\pm^\ssf{CAF} 
        &= \sqrt{K(2J+K) + \frac{B^2(J\pm J-K)}{2J+K}} \nn
        &+i\alpha \qty[K+J\pm\frac{B^2(2J\mp K)}{2(2J+K)^2}], 
        & B<B_\perp^c\\
        \omega_\pm^\ssf{FM}
        &= \sqrt{(B-J\pm J)(B-J\pm J -K)} \nn 
        &+i\alpha \qty(B-J\pm J-\frac{K}{2}),
        & B\ge B_\perp^c.
    \end{align}
\end{subequations}


\section{Appendix C: The extra damping-related phase delay between the two magnetic sublattices in AF}

The equation of motion derived from linearized coupled LLG equation for the bipartite antiferromagnet is
\begin{equation}
    -i\frac{d}{dt}
    \mqty(1-i\alpha & 0 \\ 0& 1+i\alpha)
    \Psi
    =\mqty( K+|J| & |J'|\\ -|J'| & -K-|J|)
    \Psi,
\end{equation}
where $\Psi = (\psi_1, \psi_2)^T$ with $\psi_j=m_j^x+i m_j^y$ the transverse magnetizatic component in complex form. The eigenmodes of the EOM above are
\begin{subequations}
\begin{align}
    \Psi_+ &= \mqty(-\cosh r \\ (1-i\alpha) \sinh r) e^{+i\Omega t}  
    \simeq \mqty(-\cosh r \\ e^{-i\alpha} \sinh r) e^{+i\Omega t} \\
    \Psi_- &= 
    \mqty(-\sinh r \\ (1-i\alpha) \cosh r) e^{-i\Omega t}
    \simeq \mqty(-\sinh r \\ e^{-i\alpha} \cosh r) e^{-i\Omega t},
\end{align}
\end{subequations}
with $\tanh(2r)=|J'|/(K+|J|)$. These eigenstates means that when the Gilbert damping is taken into account, there is an extra phase delay of $\alpha$ between the two magnetic sublattices, in additional to the original $\pi$ phase.


In comparison, we consider the equation of motion for the bipartite ferromagnet 
\begin{equation}
    -i\frac{d}{dt}
    (1-i\alpha) \Psi 
    = \mqty( K+|J| & -|J'| \\ -|J'| & K+|J|)
    \Psi,
\end{equation}
whose eigenmodes are
\begin{equation}
    \Psi_\pm = \frac{1}{\sqrt{2}}
    \mqty(1 \\ \pm 1) e^{i\omega_\pm^\ssf{FM}}.
\end{equation}
Therefore, the eigenstates for the FM are not modified by the Gilbert damping, different from the AF case above.


\section{Appendix D: Two-mode squeezing in AF}
The magnon Hamiltonian for bipartite antiferromagnet can be written as
\begin{equation}
    \hH=(K+|J|)(\ha_1^\dagger \ha_1+\ha_2\ha_2^\dagger)+|J'|(\ha_1^\dagger \ha_2^\dagger+\ha_1\ha_2).
\end{equation}
It can be diagonalized by Bogoliubov transformation
\begin{equation}
    \begin{pmatrix}
        \hb_1 \\ \hb_2 \\ \hb_1^\dagger \\ \hb_2^\dagger
    \end{pmatrix}=
    \begin{pmatrix}
        \cosh r &0 &0&-\sinh r \\
        0& \cosh r& -\sinh r& 0 \\
        0 &-\sinh r&\cosh r & 0 \\
        -\sinh r&0&0&\cosh r
    \end{pmatrix}\begin{pmatrix}
        \ha_1 \\ \ha_2 \\ \ha_1^\dagger \\ \ha_2^\dagger 
    \end{pmatrix},
\end{equation}
with $\tanh(2r)=J'/(K+J)$. The parameter $r$ is assumed to be real for simplicity. The diagonalized Hamiltonian is 
\begin{equation}
    \hH=\Omega(\hb_1^\dagger \hb_1+\hb_2 \hb_2^\dagger).
\end{equation}
By introducing the two-mode squeezing operator with squeezing parameter $r$ 
\begin{equation}
    U_r = e^{r(\ha_1\ha_2-\ha_1^\dagger \ha_2^\dagger)},
\end{equation}
the Bogoliubov transformation can be presented as $\hb_j = U_r^\dagger \ha_j U_r$. Therefore, the AF eigenstates are squeezed states with squeezing parameter $r$.


\end{document}